\documentclass[%
 aip,
 amsmath,amssymb,
 reprint,%
]{revtex4-1}
\usepackage{CJK}
\usepackage{graphicx}
\usepackage{color}
\usepackage{braket}
\usepackage{upgreek}
\usepackage{xr-hyper}
\makeatletter
\newcommand*{\addFileDependency}[1]{
  \typeout{(#1)}
  \@addtofilelist{#1}
  \IfFileExists{#1}{}{\typeout{No file #1.}}
}
\makeatother

\usepackage{dcolumn} 
\usepackage{bm}     
\usepackage{amsfonts}
\usepackage[bookmarks=false,colorlinks,citecolor=blue]{hyperref}
\hypersetup{colorlinks=true,linkcolor=blue,filecolor=blue,citecolor = blue, urlcolor=blue}\usepackage{amsmath}

\usepackage[version=4]{mhchem}
\usepackage{siunitx}

\newcommand{\bfl}{\begin{flushleft}}
\newcommand{\efl}{\end{flushleft}}

\usepackage{multirow}
\usepackage{array}
\newcolumntype{P}[1]{>{\centering\arraybackslash}p{#1}}

\newcommand{\ITES}{I_{\mathrm{TES}}}

\newcommand{\Ib}{I_{\mathrm{bias}}}
\newcommand{\IbDC}{I_{\mathrm{bias},{\mathrm{DC}}}}
\newcommand{\IbAC}{I_{\mathrm{bias},{\mathrm{AC}}}}
\newcommand{\Tb}{T_{\mathrm{b}}}

\newcommand{\Rs}{R_{\mathrm{s}}}
\newcommand{\Rn}{R_{\mathrm{n}}}
\newcommand{\PhiAC}{\Phi_{\mathrm{AC}}}
\newcommand{\PhiDC}{\Phi_{\mathrm{DC}}}

\newcommand{\nocontentsline}[3]{}
\newcommand\stoptoc{%
   \let\origcontentsline\addcontentsline
   \let\addcontentsline\nocontentsline
}
\newcommand\resumetoc{%
   \let\addcontentsline\origcontentsline
}

\usepackage{subcaption}   
\usepackage{caption}
\usepackage{gensymb}
\usepackage{fancyhdr}

\makeatletter
\def\@email#1#2{%
 \endgroup
 \patchcmd{\titleblock@produce}
  {\frontmatter@RRAPformat}
  {\frontmatter@RRAPformat{\produce@RRAP{*#1\href{mailto:#2}{#2}}}\frontmatter@RRAPformat}
  {}{}
}%
\makeatother

\begin{document}

\preprint{AIP/123-QED} 

\newcommand{\urss}[1]{\ensuremath{_{\mathrm{#1}}}}


\title{Imaging the Current Distribution in Superconducting Transition-Edge Sensors with Scanning SQUID Microscopy} 



\author{Samantha Walker}
\email[Email: ]{swalker@cornell.edu}
\thanks{S. W. and A. K. contributed equally to this work.
}
\affiliation{Department of Physics, Cornell University, Ithaca, NY, 14853 USA}

\author{Austin Kaczmarek}
\affiliation{Department of Physics, Cornell University, Ithaca, NY, 14853 USA}

\author{Douglas A. Bennett}
\author{Jason E. Austermann}
\author{W. Bertrand Doriese}
\author{Shannon M. Duff}
\author{Johannes Hubmayr}
\affiliation{National Institute of Standards and Technology, Boulder, CO, 80305 USA}

\author{Kelsey M. Morgan}
\affiliation{National Institute of Standards and Technology, Boulder, CO, 80305 USA}
\affiliation{Department of Physics, University of Colorado Boulder, Boulder, CO, 80302 USA}

\author{Daniel R. Schmidt}
\author{Daniel S. Swetz}
\affiliation{National Institute of Standards and Technology, Boulder, CO, 80305 USA}

\author{Joel N. Ullom}
\author{Joel Weber}
\affiliation{National Institute of Standards and Technology, Boulder, CO, 80305 USA}
\affiliation{Department of Physics, University of Colorado Boulder, Boulder, CO, 80302 USA}

\author{Michael D. Niemack}
\affiliation{Department of Physics, Cornell University, Ithaca, NY, 14853 USA}
\affiliation{Department of Astronomy, Cornell University, Ithaca, NY, 14853}

\author{Katja C. Nowack}
\affiliation{Department of Physics, Cornell University, Ithaca, NY, 14853 USA}

\date{\today}

\begin{abstract}
Transition-edge sensors (TESs) exploit the resistive transition of a superconductor to detect minute changes in temperature. While significant progress has been made in understanding the temperature- and current-dependent resistive transition in these devices, important aspects of TES behavior under bias remain unresolved. In particular, existing TES models largely neglect spatial variations in the superconducting state and current distribution arising from device geometry and proximity effects associated with the superconducting leads and normal metal features. Using scanning superconducting quantum interference device (SQUID) microscopy (SSM), we directly image local superconducting properties and transport current in two molybdenum-gold (MoAu) X-ray TES designs. We measure the local diamagnetic response to identify superconducting regions in devices voltage-biased within the transition and the Oersted field generated by the transport current to reconstruct the corresponding current distribution. We observe bias-dependent reconfigurations of the current distribution that track current-induced changes in superconducting regions, highlighting the importance of mesoscale spatial variations for TES device physics, operation, and design.
\end{abstract}

\maketitle 

A transition-edge sensor (TES) is a sensitive thermal detector that measures incident radiation through heating of a strongly temperature-dependent resistor: a superconducting film biased within its superconducting transition \cite{Irwin2005}. TESs can be operated as bolometers to measure changes in incident power \cite{IrwinHilton2005,Hubmayr2018} or as microcalorimeters capable of resolving single photons \cite{Ullom2015}, with applications spanning millimeter wavelengths \cite{Hubmayr2018} to gamma rays \cite{Bennett2012}. X-ray TES microcalorimeter arrays \cite{Gottardi2021} have been used for ground-based particle physics experiments \cite{Rothe2018,Pepe2024,Alpert2025}, high resolution spectroscopy at beamlines \cite{Doriese2017} and sounding rockets \cite{Adams2021}, and are being actively developed for future observatories \cite{Barret2018,Smith2024,Gaskin2019}.

TESs have been extensively studied and modeled over the past decade \cite{Sadleir2010,Sadleir2011,Bennett2012,Swetz2012,Bennett2013,Morgan2017}, yet important aspects of the device physics underlying the temperature- and current-dependent resistive transition when biased, which is key to their operation as sensors, remain unresolved. To date, TESs have primarily been characterized through electrical transport measurements, which are inherently global and average over the entire device \cite{Ullom2015}. Existing models of TESs, including the two-fluid model \cite{Bennett2012LTD,Bennett2013} and descriptions of TESs as proximitized weak links \cite{Sadleir2010,Sadleir2011,Vissers2013}  
neglect aspects of the device geometry and the resulting spatial variations in the superconducting state and current distribution, as well as those arising from proximity effects associated with superconducting leads and normal metal features incorporated into the design. Swetz et al. \cite{Swetz2012} modeled the influence of the self-induced magnetic field generated by the transport current to infer the current distributions in select TES geometries. However, these conclusions are indirect and motivate direct, spatially resolved measurements of both current flow and its influence on local superconducting properties in TESs. Here, we directly visualize transport current distributions and image the current-induced changes in the spatial extent of superconductivity in two molybdenum-gold (MoAu) X-ray TES microcalorimeters using scanning superconducting quantum interference device (SQUID) microscopy (SSM).

SSM uses a SQUID to detect the magnetic flux through a small pickup loop scanned close to a sample surface, providing micrometer spatial resolution \cite{Kirtley2010}. Operated in a flux-locked loop, the SQUID detects flux, here primarily due to the Oersted field generated by the bias current. Simultaneously, we drive an AC current through an on-chip field coil concentric with the pickup loop to apply a local magnetic field to the sample \cite{Huber2008} and measure the sample-induced change in the mutual inductance between the field coil and pickup loop, $\delta M$, relative to its value far from the sample \cite{kirtley2012}. For a superconductor, this change reflects its local diamagnetic response (DR), which probes the strength of superconducting screening. Both signals are averaged over time at each pixel, therefore the resulting images map the mean transport current and local superconducting response \cite{Walker2025,Kaczmarek2026}.

\begin{figure}[!t]
    \includegraphics[width=.8\columnwidth]{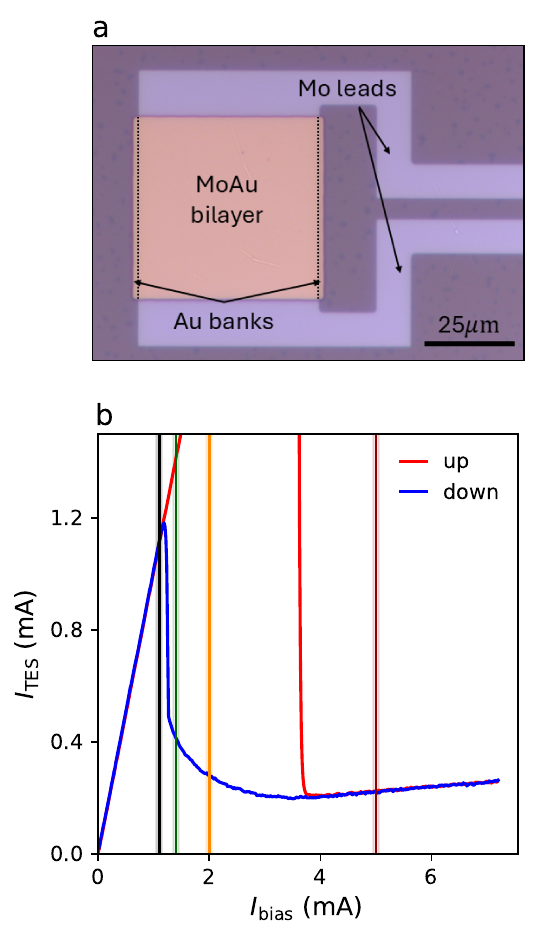}
    \caption{(a) Optical image of a $\SI{50}{\micro\meter}\times\SI{50}{\micro\meter}$ square MoAu bilayer TES with Mo leads and Au banks on a suspended SiNx membrane. (b) TES current $\ITES$ versus bias current $\Ib$ at $\Tb$\,=\,\SI{86}{\milli\kelvin} while voltage-biasing with a shunt resistor $\Rs$\,=\,\SI{180}{\micro\ohm}. Blue (red) curves show downward (upward) sweeps of $\IbDC$. Vertical lines mark DC bias points imaged in Fig.\,\ref{fig2}. Shaded regions denote the additional \SI{50}{\micro\ampere} AC current $\IbAC$ used to image how additional current distributes in the device.}
    \label{fig1}
\end{figure}

\begin{figure*}[htbp]
    \includegraphics[width=\textwidth]{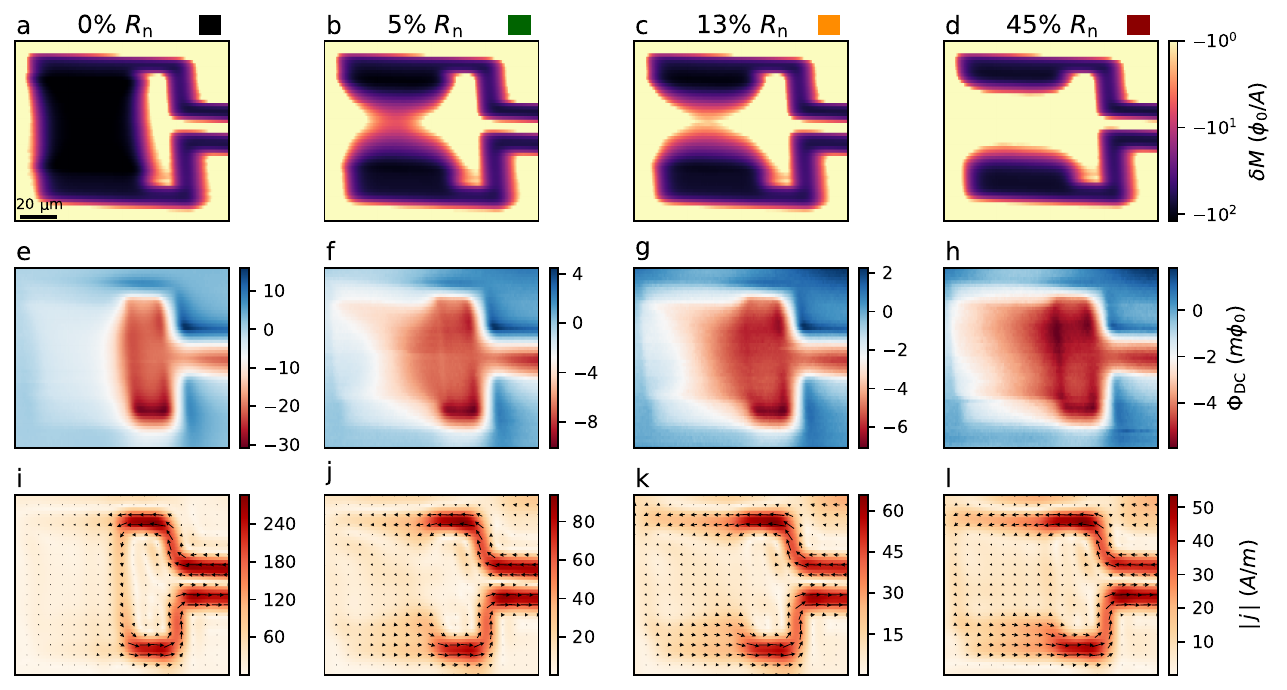}
    \caption{(a-d) Diamagnetic response (DR) of the square TES at $\Tb$ = \SI{86}{\milli\kelvin}. Colored squares indicate the bias points marked in Fig.\ref{fig1}(b) for all images in each column. From left to right, $\IbDC$ = 1.1, 1.4, 2.0, and \SI{5.0}{\milli\ampere}, corresponding to resistances of $\sim$ 0, 5, 13, and 45\,$\%\Rn$ respectively, where $\Rn$ = \SI{8.9}{\milli\ohm} is the device normal resistance. The colorbar in (d) applies to the top row. (e--h) DC-flux images of the Oersted field produced by the TES DC current 
    and (i--l) corresponding reconstructed current density. Individual color scales in (e--l) are based on measured magnitudes at each bias point, in order to highlight spatial features.
   }
    \label{fig2}
\end{figure*}

To image the TESs under operating conditions, we voltage-bias each device through a parallel shunt resistor $R\urss{s}\sim$ \SI{180}{\micro\ohm}. Instead of a typical readout using an inductively coupled SQUID and a SQUID series array amplifier \cite{Irwin2005}, we follow the approach of Walker et al. (2025) \cite{Walker2025} and position the scanning SQUID near a TES lead to measure the magnetic flux generated by the current through the TES, $\ITES$, as a function of applied DC bias current, $\IbDC$. 

We first study a $\SI{50}{\micro\meter}\times\SI{50}{\micro\meter}$ square TES shown in Fig.\ref{fig1}(a). A \SI{46}{\nano\meter} layer of Mo with a superconducting transition temperature $T_{c_L}$\,$\sim$\,\SI{0.9}{\kelvin} is first deposited in the shape of the leads and the TES square. Subsequently, a \SI{507}{\nano\meter} layer of Au is deposited on top to form a MoAu bilayer in the TES device region \cite{Weber2020}. The Au extends beyond the Mo layer to prevent shorts through non-proximitized Mo, forming \SI{1}{\micro\meter} wide normal metal banks along the MoAu bilayer edges. The bilayer has a transport transition temperature $T_c$\,$\sim$\,\SI{98.5}{\milli\kelvin}. The TES is suspended on a thin, millimeter-scale silicon nitride (SiN$_\mathrm{x}$) membrane to thermally decouple it from the cryostat.

Fig.\,\ref{fig1}(b) shows the measured $\ITES$-$\Ib$ curves for increasing and decreasing bias current at a bath temperature $T\urss{b}$\,=\,\SI{86}{\milli\kelvin}. The TES switches from the superconducting to the normal branch at $\IbDC$\,$\sim$\,\SI{3.7}{\milli\ampere} as the bias current increases and returns to the superconducting state at $\IbDC$\,$\sim$\,\SI{1.3}{\milli\ampere} as it decreases. This large hysteresis arises primarily from the thermal decoupling of the device from the bath by the SiN$_\mathrm{x}$ membrane. We image the TES at selected bias points as the bias current is reduced, bringing it into the transition from the normal branch as in typical TES operation \cite{Irwin1995}.

Fig.\,\ref{fig2} shows SSM images of the square TES at $T\urss{b}$\,=\,\SI{86}{\milli\kelvin} for the bias points marked in Fig.\ref{fig1}(b), acquired as the bias current was reduced. These span the superconducting branch (0\,$\%\Rn$, $\IbDC$ = \SI{1.1}{\milli\ampere}) and the transition ($\sim$\,5, 13, and 45\,$\%\Rn$ at $\IbDC$ = 1.4, 2.0, and \SI{5.0}{\milli\ampere}, respectively), where $\Rn$ = \SI{8.9}{\milli\ohm} is the TES normal resistance. Figs.\,\ref{fig2}(a--d) show $\delta M$ on a logarithmic scale to highlight superconducting regions; the images on a linear scale are shown in Supplement Section S3. These images reveal where superconductivity persists and how current affects the its screening response at a fixed bath temperature below $T_c$. At 0\,$\%\Rn$ (a), the TES is superconducting throughout its interior, as indicated by a strong DR across the entire bilayer. Within the transition (b--d), the DR is concentrated near the Mo leads, which remain superconducting at all bias points. At 45$\%\Rn$, the current largely suppresses the DR except near the upper and lower leads, where it remains enhanced relative to the DR of the Mo leads alone (see \SI{211}{\milli\kelvin} image in Fig. S9).  
As the bias is reduced, weak halos of DR extend farther into the bilayer as the current is decreased, while the response near the device edges remains suppressed. This spatial structure reflects the competing direct proximity effect from the superconducting Mo leads and the inverse proximity effect from the Au banks, studied in detail in \cite{Kaczmarek2026}.

The simultaneously acquired DC-flux images,  $\Phi_{\mathrm{DC}}$, in Figs.\,\ref{fig2}(e--h) show the Oersted field produced by the TES current. Figs.\,\ref{fig2}(i--l) show the corresponding reconstructed current density magnitude, $|j|$ \cite{clement2019, ferguson2023}, with arrows indicating current direction.
We note that these DC-flux images include a small background signal contribution in the absence of any current; however, these signals are much smaller in magnitude than the flux associated with the Oersted field generated by the current. From left to right, the flux and corresponding current density magnitude decrease, consistent with the splitting ratio $\ITES/\Ib$ in Fig.\ref{fig1}(b). Current follows the Mo leads and enters and exits the TES at the upper and lower right lead contacts, respectively. At 0\,$\%\Rn$ (e, i), most current within the TES follows the right edge between the contacts, while a smaller fraction flows along the leads and across the bilayer. Magnetic self-screening enhances current near superconducting edges described by the London equations \cite{Tinkham2004}. Numerical solutions of the London equations qualitatively reproduce this distribution, with some dependence on the assumed London penetration depth (Supplement Section S7). 
Within the transition (j--l), current spreads along the upper and lower Mo leads and flows across the bilayer, becoming more uniform with increasing $\%\,\Rn$. At lower resistance, the flux images suggest an hourglass-like current distribution correlated with the DR, though this structure is less apparent in the reconstructed current density because of a limited signal-to-noise ratio. Simultaneous imaging of the flux in response to a small AC modulation of the transport bias, reveals a more clear correlation: a small additional current preferentially follows the hourglass-shaped regions that retain superconducting screening, even at 13\,$\%\Rn$ (see Supplement Section S4). Both the current and DR images indicate that proximity effects widen the superconducting leads beyond the Mo strips, effectively reducing the normal region of the device even at 45\,$\%\Rn$. Comparing with an image at $\Tb$ = \SI{211}{\milli\kelvin}, where proximity effects are suppressed, confirms that the DR near the leads remains enhanced (Supplement Section S6).

At 5\,$\%\Rn$, the DR image (Fig.\,\ref{fig2}(b)) suggests there is a continuous superconducting path through the center of the hourglass, yet transport shows finite resistance (Fig.\ref{fig1}(b)). Dissipation could arise from vortex-mediated flux flow, as considered previously \cite{Bulaevskii2011,Ezaki2018,Fàbrega2024,Bennett2014}, or from a small normal region near the TES center below our spatial resolution. Measurements with high temporal resolution combined with modeling that captures phase-slip dynamics could help distinguish between these possibilities by resolving vortex motion across the hourglass pinch point.

\begin{figure}[t]
  \centering
    \includegraphics[width=.8\columnwidth]{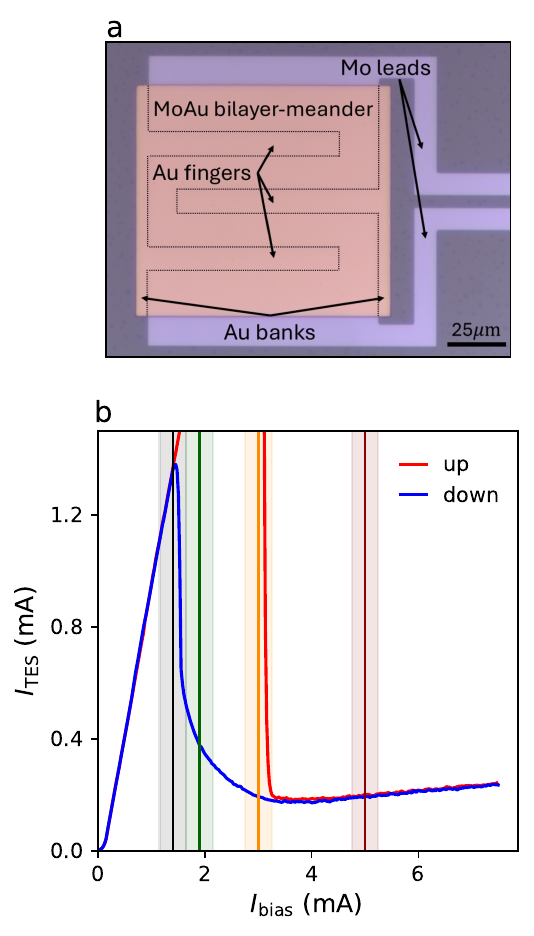}
    \caption{(a) Optical image of the $100\times\SI{100}{\micro\meter}$ TES with Mo leads, a MoAu bilayer meander, Au fingers, and Au banks on a suspended SiNx membrane. (b) TES current $\ITES$ versus bias current $\Ib$, where the TES is voltage-biased with a shunt resistor $\Rs$ = \SI{180}{\micro\ohm} at $\Tb$ = \SI{62}{\milli\kelvin}. Blue (red) curves show to downward (upward) sweeps of $\IbDC$. Vertical lines mark DC bias points $\IbDC$ imaged in Fig.\ref{fig4}. Shaded regions indicate the additional \SI{250}{\micro\ampere} AC bias $\IbAC$ used to image how additional current distributes in the device.}
    \label{fig3}
\end{figure}

\begin{figure*}[htbp]
  \centering
    \includegraphics[width=\textwidth]{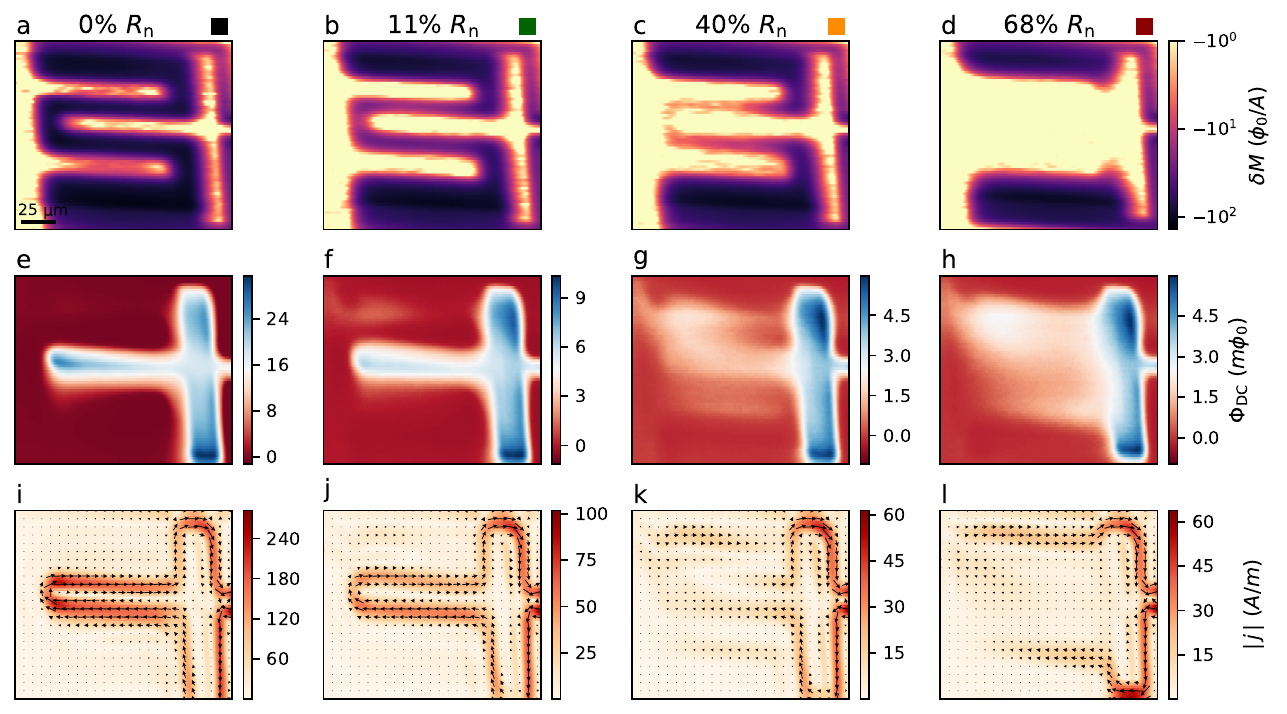}
    \caption{(a-d) Diamagnetic response of the meander TES at $\Tb$ = \SI{62}{\milli\kelvin}. Colored squares identify the bias points marked in Fig.\ref{fig3}(b) for all images in each column. From left to right, $\IbDC$ = 1.4, 1.9, 3.0, and \SI{5.0}{\milli\ampere}, corresponding to resistances of $\sim$ 0, 11, 40, and 68\,$\%\Rn$, respectively, where $\Rn$ = \SI{6.4}{\milli\ohm} is the normal state resistance. The colorbar in (d) applies to the top row. (e--h) DC-flux images of the Oersted field produced by the DC current at each bias point (a--d) and (i--l) corresponding reconstructed current density. Individual color bars in (e--l) are based on measured magnitudes, in order to highlight spatial features.
    }
    \label{fig4}
\end{figure*}

We next consider a TES whose more complex geometry produces richer spatial structure in the DR and current distribution (Fig.\,\ref{fig3}(a)). As in the square TES, a \SI{46}{\nano\meter} layer of Mo is deposited first, but here it  is patterned into a meander within the TES region. A \SI{507}{\nano\meter} Au layer is then deposited over a $\SI{100}{\micro\meter}\times\SI{100}{\micro\meter}$ area, forming a MoAu bilayer meander, \SI{5}{\micro\meter} wide Au banks along the edges, and \SI{10}{\micro\meter}-wide Au fingers without underlying Mo on the TES. The transport transition temperature is $T_c$\,$\sim$\,\SI{78.5}{\milli\kelvin}, lower than $\sim$\,\SI{98.5}{\milli\kelvin} for the square device. This reduction arises from the additional inverse proximity effect from the Au fingers and the current distribution in the meander.

Fig.\,\ref{fig3}(b) shows $\ITES$-$\Ib$ curves for the meander TES at $\Tb$\,=\,\SI{62}{\milli\kelvin} for sweeps of $\IbDC$. Despite the lower bath temperature and larger device area, which would nominally reduce the current density, the critical current of $\sim$\,\SI{3.1}{\milli\ampere} on the upward sweep is lower than that of the square device.

Fig.\,\ref{fig4} shows SSM images of the DR (a--d), DC-flux (e--h), and reconstructed current density (i--l) at $\Tb$ = \SI{62}{\milli\kelvin}. The DR is shown on a logarithmic scale; linear-scale images are available in Fig. S4. The images were acquired as the bias current was reduced, spanning the superconducting branch (0\%\,$\Rn$, $\IbDC$ = \SI{1.4}{\milli\ampere}) and the transition at 11, 40, and 68\,$\%\Rn$ ($\IbDC$ = 1.9, 3.0, and \SI{5.0}{\milli\ampere}, respectively), where $\Rn$ = \SI{6.4}{\milli\ohm} is the device normal resistance. At 0\,\%$\Rn$ (a), the DR is pronounced  and relatively uniform along the meander, with a faint response in the Au fingers, suggesting proximity-induced superconductivity in the fingers. At 11\,\%$\Rn$ (b), a continuous superconducting path remains visible along the meander, while at 40\,\%$\Rn$ (c), the DR becomes much weaker, particularly along the leftmost segment. At 68\,\%$\Rn$ (d), the TES is largely normal and a DR is only observed in the Mo leads and in the adjacent bilayer, extending slightly into the rightmost vertical segments of the meander. As in the square device, finite resistance coexists with an apparently continuous superconducting path in (b--c). This could reflect unresolved normal regions or dissipation from vortex motion, particularly in segments with weak DR as discussed above.

The current distributions (Figs.\,\ref{fig4}(i--l)) obtained from  DC-flux images $\Phi_{DC}$ of the Oersted field generated by the current (Figs.\,\ref{fig4}(e--h)) evolve alongside these changes in DR. 
At 0\%\,$\Rn$ (i), current is primarily confined to the bilayer meander. However, imaging the response to a small AC modulation of the bias shows that additional current crosses the bottom Au finger along a faint diagonal path (Supplement Section S4), consistent with the weak DR in portions of the Au fingers that suggest that they become superconducting (a). At 11\,$\%\Rn$ (j), current follows the meander almost entirely, with only the imaging in response to a small AC modulation revealing that a fraction flows in parts of the proximitized bilayer near the Mo leads (Supplement Section S4). At 40 $\%\Rn$ (k), current appears split between the bilayer meander, through the top and bottom Mo leads, and into proximitized, adjacent bilayer regions. The response to the AC modulation of the bias shows that additional current preferentially follows the meander, with a smaller fraction crossing between its segments. This is consistent with (c) where the leftmost meander segment does not appear to be fully superconducting, making the current distribute between alternative resistive paths.
At 68 $\%\Rn$ (l), the device is now mostly driven normal by the current (d) and current spreads relatively uniformly throughout the device. Similar to the square device, the bilayer adjacent to the Mo leads remains superconducting effectively (see Supplement Section S6 for images at elevated temperatures) shortening the length of the normal region between the two leads.

\begin{figure*}[htbp]
  \centering
    \includegraphics[width=\textwidth]{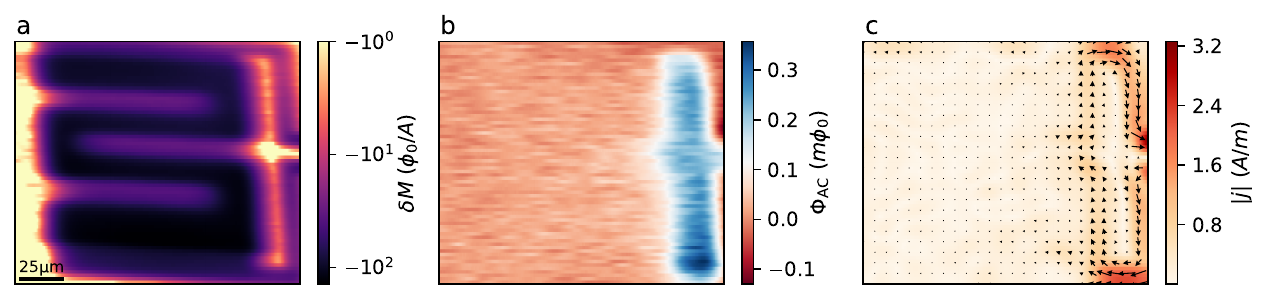}
    \caption{(a) Diamagnetic response of the meander TES with $\IbAC$ = \SI{20}{\micro\ampere} and zero DC bias at $\Tb$ = \SI{20}{\milli\kelvin}. Unlike within the transition (Fig.\,4), a finite diamagnetic response appears within the Au fingers.  
    (b) AC-flux image of the Oersted field produced by $\IbAC$ and (c) corresponding reconstructed AC current density.}
    \label{fig5}
\end{figure*}

Finally, we image the meander device at a lower bath temperature ($\Tb$=\SI{20}{\milli\kelvin}) and low current ($\IbDC=0$, $\IbAC$\,=\,\SI{20}{\micro\ampere}) to study magnetic self-screening when the Au fingers become superconducting. A DR within the Au fingers indicates proximity-induced superconductivity (Fig.\,\ref{fig5}(a)). The AC-flux image (b) and reconstructed current density in (c) show current concentrated along the right edge, resembling the square TES in the superconducting state (Figs.\,\ref{fig2}(e) and (i)) and indicating that self-screening shapes the current distribution. This demonstrates that the MoAu bilayer can proximitize Au traces as wide as \SI{10}{\micro\meter} sufficiently to support superconducting screening currents.

Unlike in the square TES, a small indention in (Fig.\,\ref{fig5}(b)) indicates that current shifts inward near the central Au finger (c), consistent with locally suppressed DR due to the inverse proximity effect from the right Au bank (a). The current redistribution between Fig.\,\ref{fig4}(i) to Fig.\,\ref{fig5}(c) occurs while in in the zero-resistance state and is therefore difficult to access via transport measurements \cite{Swetz2012}.

Many TES properties, including speed and energy resolution, depend on the partial derivatives $\alpha = \partial \log R/\partial \log T$ and $\beta = \partial \log R/\partial \log I$ \cite{IrwinHilton2005}. Our measurements reveal mesoscale spatial structure in both superconducting regions and transport current in TESs, including self-screening effects, which may influence these parameters. For example, in the square device (Fig.\,\ref{fig2}), the lead geometry combined with London screening concentrate current near the device edge closest to the contacts. Different lead configurations, such as leads connected at the TES midline rather than along one edge, could therefore modify the current distribution and device behavior. In the meander device, the Au fingers become superconducting through proximity effects at low temperatures (Fig.\,\ref{fig5}), despite being $\SI{10}{\micro\meter}$ wide. It is therefore reasonable to expect that devices with wider fingers could exhibit qualitatively different behavior.

In summary, we have directly visualized local superconducting properties and transport current distributions in MoAu X-ray TES microcalorimeters tuned through the superconducting transition by varying the applied DC bias. At fixed bath temperatures below $T_c$, DR imaging reveals where the applied current weakens or completely suppresses superconductivity. The current distributions are strongly influenced by self-screening effects and the spatial structure of superconducting regions, which is shaped by proximity effects from superconducting leads and normal metal structures incorporated into the device design \cite{Kaczmarek2026}. At some bias points in both devices, the DR suggests a continuous superconducting path connecting the leads, yet the device remains resistive. Clarifying whether this resistance arises from normal regions below our spatial resolution or dissipation within weak superconducting regions requires further investigation. Measurements with higher temporal and spatial resolution combined with modeling that captures phase-slip dynamics could help distinguish these possibilities. 
Our results underscore the importance of mesoscale spatial variations in the superconducting state and current distribution, including self-screening, in TESs. Incorporating spatially-dependent behavior into future TES models may therefore be important to better understand and predict device performance.

\stoptoc
\section*{Supplementary Material}
See the Supplementary Material for additional details, including more information about the devices, the scanning SQUID and experimental methods, diamagnetic response images plotted on a linear scale, flux images in response to an additional AC current, more details about the current reconstruction, images of the current distribution as a function of temperature, and London modeling for the square device.

\begin{acknowledgments}
This work was supported by the Air Force Research Laboratory, Project Grant FA9550-21-1-0429. S. Walker acknowledges support by the National Science Foundation Award No. 2503181 and the Cornell Center for Materials Research.
\end{acknowledgments}

\section*{Author Declarations}
\subsection*{Conflict of Interest}
The authors have no conflicts to disclose.

\section*{Data Availability}
The data that supports the findings of this study are available from the corresponding author upon reasonable request.


%

\resumetoc
\clearpage

\pagebreak
\widetext
\begin{center}
\textbf{\Large Supplementary Material for: Imaging the Current Distribution in Superconducting Transition-Edge Sensors with
Scanning SQUID Microscopy}
\end{center}
\setcounter{equation}{0}
\setcounter{figure}{0}
\setcounter{table}{0}
\setcounter{page}{1}
\makeatletter

\renewcommand{\theequation}{S\arabic{equation}}
\renewcommand{\thesection}{S\arabic{section}}  
\renewcommand{\thepage}{S\arabic{page}} 
\renewcommand{\thefigure}{S\arabic{figure}}
\renewcommand{\thetable}{S\arabic{table}}



\hypersetup{linkcolor=red} 
\tableofcontents

\newcounter{somecounter}


\newpage
\stepcounter{somecounter}

\section{Supplementary Section \thesomecounter: Additional Device Details}

\begin{figure}[htbp]
    \includegraphics[width=.8\linewidth]{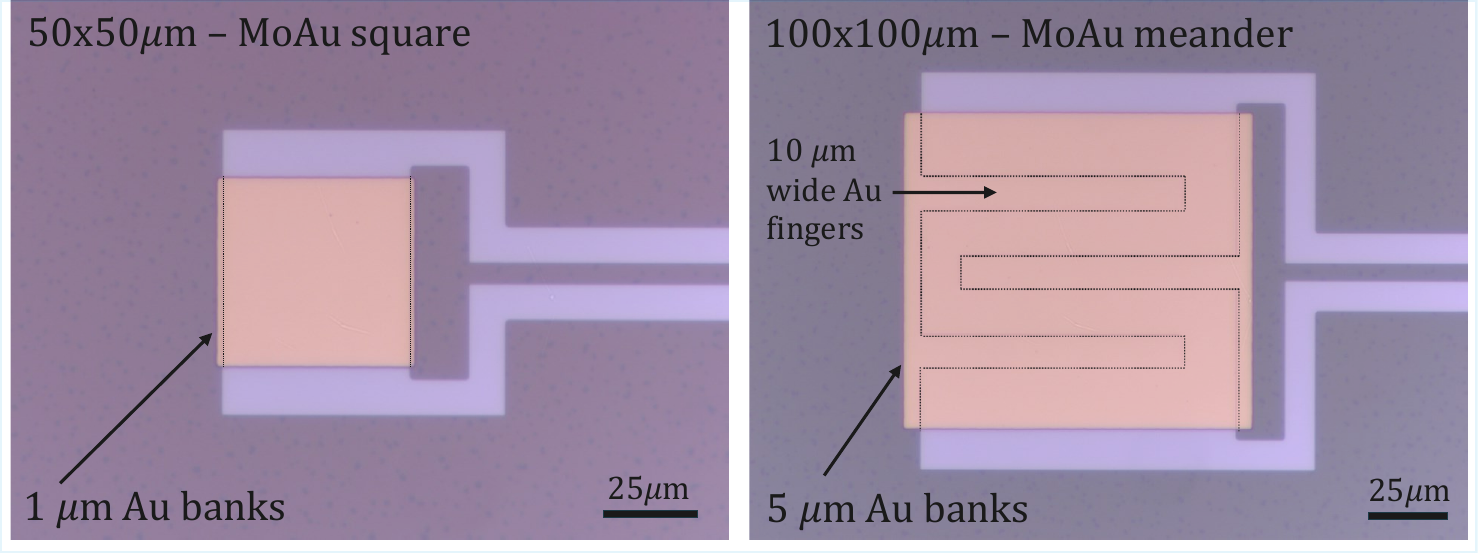}
    \caption{Optical images of the two molydenum gold (MoAu) transition-edge sensor (TES) device geometries studied in this work with additional device details. These are prototype devices for X-ray microcalorimetry applications. Here, \SI{46}{\nano\meter} of sputtered Mo is patterned into the shape of the leads and either a square or meandering shape in the TES device region. \SI{507}{\nano\meter} of Au was evaporated on top of the TES device region, forming a MoAu bilayer. The gold layer extends beyond the TES device region, forming Au only banks along the device edges with a $\SI{1}{\micro\meter}$ width for the square device and a $\SI{5}{\micro\meter}$ width for the meander device. In the \SI{100}{\micro\meter} meander device, the Au layer additionally lies in the cutouts of the underlying Mo meander, resulting in Au only fingers. The silicon (Si) is back etched, suspending TES on a \SI{2}{\micro\meter} thick silicon nitride (SiN$_x$) membrane \cite{Weber2020}. Both TESs are proximitized superconducting systems. We measure the onset of a diamagnetic response using the scanning SQUID, on a large-area MoAu device on the same chip to be $T_{c_0}\sim\SI{92.8}{\milli\kelvin}$ reflecting the intrinsic critical temperature of the bilayer \cite{Kaczmarek2026}.}
    \label{fig:all_optical_imgs}
\end{figure}

\clearpage
\newpage

\stepcounter{somecounter}
\section{Supplementary Section \thesomecounter: Scanning SQUID}\label{sec:SQUID}

\begin{figure}[htbp]
    \includegraphics[width=0.6\linewidth]{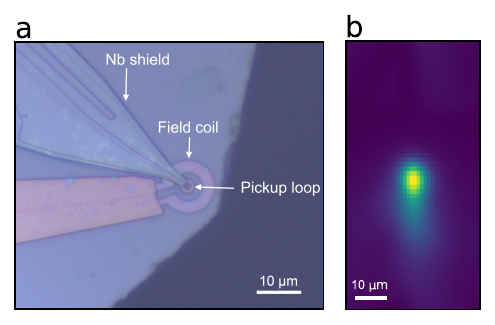}
    \caption{(a) Optical image of the scanning SQUID susceptometer used in this work, with important components labeled. The pickup loop has an inner and outer radius of $\SI{0.75}{\micro\meter}$ and $\SI{1.6}{\micro\meter}$, respectively; and the field coil has an inner and outer radius of $\SI{3.0}{\micro\meter}$ and $\SI{6.0}{\micro\meter}$, respectively.
    (b) DC-flux image $\PhiDC$ of a single vortex in a 2 mm thick superconducting sample revealing  an elongated teardrop-like shaped point spread function (PSF) of the scanning SQUID. The tail in the teardrop is due to a short section of parallel leads to the pickup loop that is only partially shielded by the Nb shield. Using the known stray magnetic field of an Abrikosov vortex, we numerically extract the PSF, fit it with a multi-Gaussian model to this PSF, and use this model as PSF in the current reconstruction.}
        \label{fig:SQUID}
\end{figure}

The scanning superconducting quantum interference device (SQUID) microscope (SSM) is installed in a dilution refrigerator \cite{Low2021} to enable imaging at milliKelvin temperatures, where TESs are typically operated. Fig.\,S2(a) shows an optical image of the scanning SQUID used in this work, where the Nb shield, field coil, and pickup loop of the SQUID are important components. 

To measure the local diamagnetic response (DR), we apply an AC current through the field coil to apply a local magnetic field to the sample. For the data presented in this work, we apply an AC current of \SI{86}{\micro\A} at \SI{277.7}{\hertz} through the field coil. The pickup loop, which is connected to a SQUID circuit, detects the resulting local magnetic field. Far from the sample, we measure the bare mutual inductance $M_0$ between the field coil and the pickup loop by measuring the flux in the SQUID at the applied frequency using a lock-in amplifier. When the SQUID is brought near the surface of a superconductor, screening currents in the sample are induced over the length scale set by the field coil geometry. These screening currents reduce the local field coupled into the pickup loop, resulting in a reduction in the mutual inductance between the two loops, $\delta M$, and allow us to visualize superconducting regions/the diamagnetic response of a device as we scan over the sample surface \cite{Walker2025,Kaczmarek2026}. 

In a sample with uniform superfluid stiffness, $\delta M$ can be modeled analytically \cite{kirtley2012} in terms of the sensor geometry and the superfluid stiffness (or equivalently penetration depth $\lambda$) of the sample. Near sample boundaries or in a spatially non-uniform superconductor, quantitatively interpreting the measured $\delta M$ becomes more complicated. In these cases, the screening currents in response to the field applied by the field coil flow through regions with varying superfluid stiffness in the superconductor or may be distorted due to a sample boundary. Although numerical modeling of the DR is possible in principle \cite{BISHOPVANHORN2022}, it requires detailed assumptions about the spatial profile the superconducting properties and is therefore challenging. 

For the purpose of this work, the most significant effect is the finite spatial resolution with which we can resolve changes in the superfluid density. The measured signal at one pixel location has contributions from screening currents flowing in nearby regions. For example, if the pickup loop is centered over a non-superconducting region, or a region with much weaker superfluid stiffness, the measured signal may be enhanced by adjacent regions with stronger screening.

To image the current distributions, we apply a DC bias current to the voltage biasing circuit that sets the operating point of the resistively shunted TES. We then image the DC flux produced by the current flowing in the TES. In addition, we add a small AC current oscillating at \SI{23}{\hertz} to the DC bias current, and measure AC flux coupled through the SQUID pickup loop at this frequency simultaneously using a lock-in amplifier and recording the signal on a separate analog-to-digital converter (ADC) channel. Another lock-in amplifier measures the diamagnetic response as described detecting at \SI{277.7}{\hertz}. This signal is recorded on a separate analog-to-digital converter (ADC) channel as well. Both signals are averaged over time at each pixel, where we used a time constant of 0.3 s and a 0.3--1 Hz filter to set a scan rate that would balance the signal-to-noise and sampling time spent on each pixel. The resulting images map the mean transport current and local superconducting response. 

As shown in Fig.\,S2(b), the pickup loop geometry and height of the SQUID above the sample result in an imaging point spread function (PSF) of the SQUID (see also Supplement Section \ref{sec:reconstruction}). The PSF for our SQUID has an elongated teardrop-like shape, whose influence contributes to the flux images. Fig.\,S2(b) shows a DC-flux image $\PhiDC$ of a single vortex in a 2 mm thick superconducting film where this elongated teardrop-like shape is clearly visible. Using the known stray magnetic field of an Abrikosov vortex, we numerically extract the PSF, fit it with a multi-Gaussian model to this PSF, and use this model as PSF in the current reconstruction to minimize distortions in the reconstructed current density that might arise due to this PSF.

\newpage
\newpage
\newpage

\stepcounter{somecounter}
\section{Supplementary Section \thesomecounter: Diamagnetic Response Images on a Linear Scale}
\begin{figure}[htb]
    \includegraphics[width=\columnwidth]{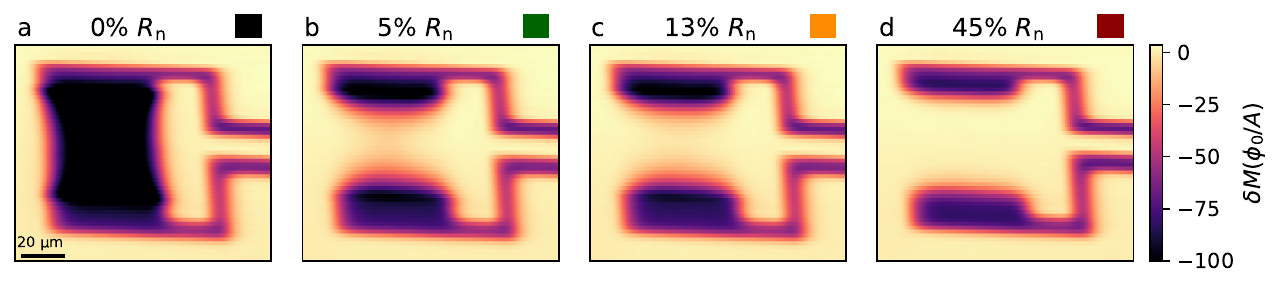}
    \caption{Imaging of the diamagnetic response in the square TES (same as Fig.2(a-d) in the main text), plotted on a linear color scale.
    }
    \label{fig:square_lin_scale}
\end{figure}

\begin{figure}[htb]
    \includegraphics[width=\columnwidth]{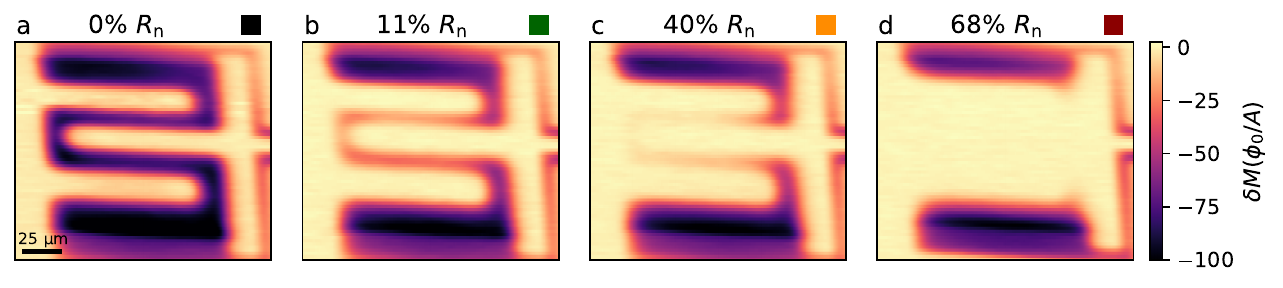}
    \caption{Imaging of the diamagnetic response in the meander TES (same as Fig.4(a-d) in the main text), plotted on a linear color scale.
    }
    \label{fig:meander_lin_scale}
\end{figure}

\newpage
\clearpage
\newpage

\stepcounter{somecounter}
\section{Supplementary Section \thesomecounter: Flux Images in Response to an Additional AC Current}

In the main text, we present and discuss flux images of the Oersted field produced by a DC bias current $\IbDC$, $\Phi_{\mathrm{DC}}$, in TESs at different bias points within their transitions and the corresponding current distributions. In addition to $\IbDC$, at each bias point, we also applied a small additional AC bias current $\IbAC$, $\Phi_{\mathrm{AC}}$, oscillating at 23 Hz, to measure where a small amount of additional current would preferentially flow. Using lock-in detection, we measure the magnetic flux $\Phi_{\mathrm{AC}}$ associated with the Oersted field generated by $\IbAC$ simultaneously with $\Phi_{\mathrm{DC}}$ measurements on separate ADC channels. Lock-in detection eliminates contributions that may couple into the DC-flux channel, such as small background magnetic field gradients. For the square TES, we apply an additional $\IbAC$\,=\,\SI{50}{\micro\ampere} peak-to-peak and for the meander device, we applied an additional $\IbAC$\,=\,\SI{250}{\micro\ampere} peak-to-peak. These $\IbAC$ are indicated by the width of shaded regions around each vertical line/bias point on the down-sweep $\ITES$-$\Ib$ curves in Fig.\,1(b) and Fig.\,3(b) in the main text, respectively. For the meander device, we use a larger $\IbAC$ to improve the signal-to-noise of the measured flux and to be able to image at higher bias, where the fraction of $\IbAC$ that reaches the TES and the corresponding flux signal is reduced.

The resulting AC-flux images $\Phi_{\mathrm{AC}}$ for the square and meander devices for the same bias points as Figs.\,2 and 4 in the main text are shown in Figs.\,S5(a--d) and Figs.\,S6(a--d). The corresponding AC current density magnitudes are shown in Figs.\,S5(e--h) and Figs.\,S6(e--h). These images probe the time-averaged change in the current distribution and are therefore sensitive to changes in the magnitude, phase, and the direction of current flow in the device. More generally, images acquired with lock-in detection reflect the behavior of the device averaged over the range sampled by the AC excitation. Color bars differ for each image, where magnitudes are based on measured raw data values at each bias point, in order to clearly show spatial features in the data. While we discuss these measurements in the main text, we note some general features here. The fraction of $\IbAC$ that flows through the TES at each bias point is determined by the splitting ratio $\partial \ITES/\partial \Ib$ at the operating point set by $\IbDC$. From (a--d), in Figs.\,S5 and S6, the measured signal magnitudes in AC-flux images decrease as the splitting ratio decreases. Similarly, the corresponding reconstructed current density magnitudes also decrease.

For the square device, Fig.\,S5(a) shows a flip in the the color map relative to (b--d) due to a phase reversal in the AC current when the TES returns to the superconducting branch compared to when the TES is biased on a down-sweep of $\IbDC$.The reconstructed current densities show a corresponding reversal in the current direction as indicated by the arrows. This reversal arises from the change in the slope of the $\ITES$-$\Ib$ curves (i.e. a sign change in  $\partial \ITES/\partial \Ib$) between the two regimes: in the normal branch, the negative slope means that when $\IbAC$ increases (decreases), $\ITES$ decreases (increases), whereas in the superconducting branch, the positive slope means $\ITES$ increases (decreases) as $\IbAC$ increases (decreases). The sign of the AC-flux color map reverses again between (c) and (d) due to the change in slope of the down-sweep $I_{\mathrm{TES}}$–$I_{\mathrm{bias}}$ curve from negative to positive. The meander device shows similar flips in the color map in Fig.\,S6(a) relative to (b--c) and reverses again between (c) and (d) due to the change in the slope of the $\ITES$-$\Ib$ curves from positive to negative and negative to positive.
\\
\begin{figure}[htb]
    \includegraphics[width=\columnwidth]{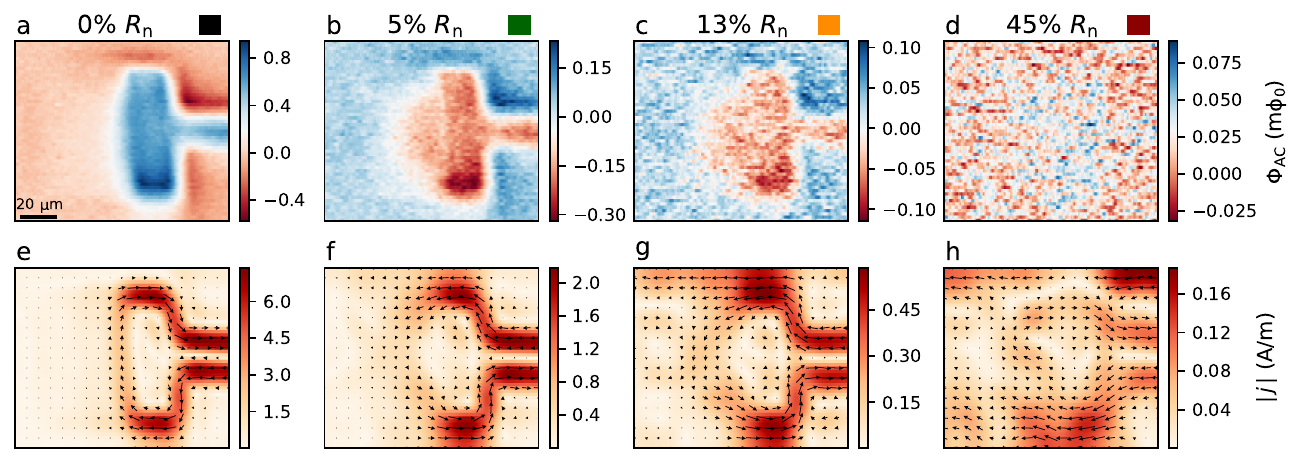}
    \caption{AC-flux images $\PhiAC$ of the Oersted field in the square TES produced by a \SI{50}{\micro\ampere} AC bias current $\IbAC$ peak-to-peak at each bias point. Colored squares and \%$\Rn$ labels correspond to $\IbDC$ bias points indicated by the vertical dashed lines in Fig.1(b) in the main text. Images were measured simultaneously on a separate ADC channel from DC-flux images shown in Fig.2(e-h) in the main text. Color bars differ for each image, where magnitudes are based on measured raw data values at each bias point, in order to clearly show features in the data. The 45\,\%$\Rn$ AC-flux image and corresponding current density are mostly noise compared to the $\PhiDC$ data since only a tiny fraction of the \SI{50}{\micro\ampere} AC bias current flows across the device at this bias point.
    }
    \label{fig:square_phi_DC}
\end{figure}

\begin{figure}
    \includegraphics[width=\columnwidth]{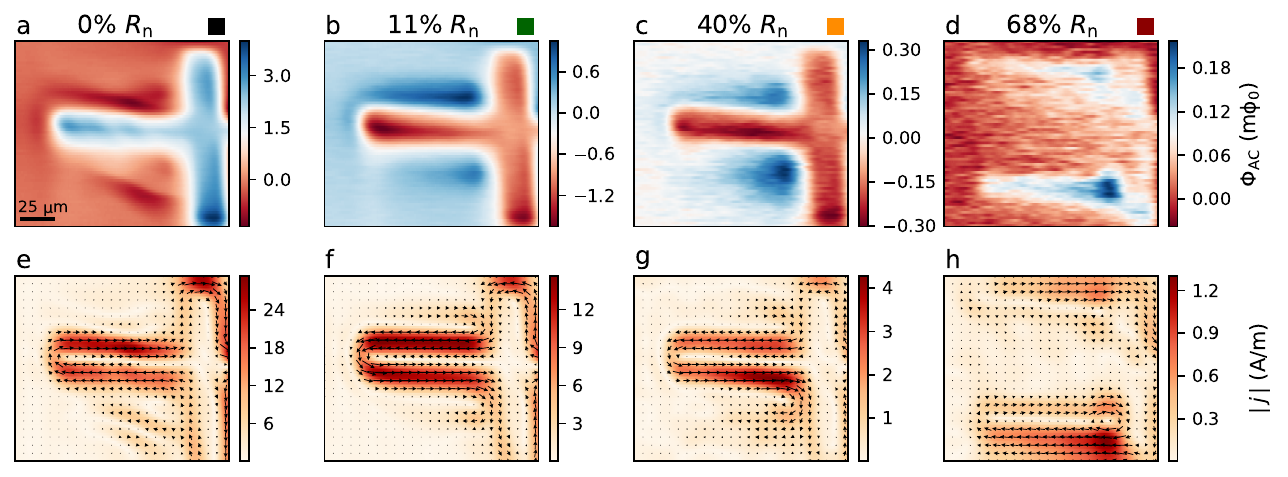}
    \caption{AC-flux images $\PhiAC$ of the Oersted field in the meander TES produced by a \SI{250}{\micro\ampere} $\IbAC$ peak-to-peak at each bias point. Colored squares and \%$\Rn$ labels correspond to $\IbDC$ bias points indicated by the vertical dashed lines in Fig.3(b) in the main text. Images were measured simultaneously on a separate ADC channel from DC-flux images shown in Fig.4(e-h) in the main text. Color bars differ for each image, where magnitudes are based on measured raw data values at each bias point, in order to clearly show features in the data.}
    \label{fig:meander_phi_DC}
\end{figure}

\newpage
\clearpage
\newpage

\stepcounter{somecounter}
\section{Supplementary Section \thesomecounter: Current Reconstruction}\label{sec:reconstruction}
The current reconstruction method detailed below is adapted from Ferguson et al. (2023)\cite{ferguson2023}. The magnetic flux image measured by the SQUID $\Phi(x,y)$ at fixed height $z$ above the device surface from current flowing in the sample is a convolution of the SQUID point spread function $K_\mathrm{PSF}(x,y)$ to account for the pickup loop geometry, a Biot-Savart kernel $K_\mathrm{BS}(x,y)$, and the current stream function $g(x,y)$, such that \cite{clement2019,ferguson2023}:
\begin{equation}
    \Phi(x,y) = K_\mathrm{PSF}(x,y) * K_\mathrm{BS}(x,y) * g(x,y).
\label{eq: measuredflux_0}
\end{equation}
where $* $ denotes a convolution. The goal of the reconstruction is then to determine $g(x,y)$ from the flux image $\Phi(x,y)$ by deconvolving Eq.\ref{eq: measuredflux_0}. This is in principle possible if we assume the current does not vary through the thickness of the sample, and the 2D current density is then related to $g(x,y)$ by $\boldsymbol{j} = \nabla \times g(x,y) \hat{z}$. We combine $K_\mathrm{PSF}(x,y)$ and $K_\mathrm{BS}(x,y)$ into a single linear matrix operator $M$ for the imaging kernel, such that $\boldsymbol{\Phi} = M\boldsymbol{g}$. To construct $M$, we use the Biot-Savart kernel $K_\mathrm{BS}(x,y)$ for a dipole field associated with a current segment located at each pixel along with a previously measured $K_\mathrm{PSF}(x,y)$ for the SQUID sensor (see Fig. S2). Since the Biot-Savart kernel is long-ranged, the inverse operator $M^{-1}$ contains high spatial frequencies, and when operating on a flux image with any amount of noise, as in a real experiment, the noise is amplified, and in particular, there is a large space of $\boldsymbol{g}$ that can perfectly fit the data including the noise, but which correspond to non-physical solutions with very quickly changing current densities from pixel to pixel.We therefore regularize the inversion to penalize these solutions and favor smoother, more physically meaningful ones. The problem is therefore cast as finding a $\boldsymbol{g^*}$ which minimizes the square of the residuals between the measured flux image and the reconstructed image with an additional cost function $l(\boldsymbol{g})$ to penalize non-physical solutions:
\begin{equation}
\label{eq: measuredflux}
    \boldsymbol{g^*} = min_{\boldsymbol{g}}(||M\boldsymbol{g}-\boldsymbol{\Phi}||^2 + \sigma l(\boldsymbol{g})),
\end{equation}
Here, $l(\boldsymbol{g})$ is the function that regularizes $\boldsymbol{g}$ and $\sigma$ sets the strength of the regularization. The importance of $l(\boldsymbol{g})$ is to penalize quickly varying, non-physical solutions of $\boldsymbol{g}$, and therefore $l(\boldsymbol{g})$ typically involves derivatives of $\boldsymbol{g}$, favoring more smooth solutions of $\boldsymbol{g}$ and therefore $\boldsymbol{j}$. Here, we choose to penalize all second derivatives in $\boldsymbol{g}$ and therefore for first derivatives in $\boldsymbol{j}$, with $l(g)$ taking the form:

\begin{equation}
\label{eq: measuredflux}
    l(g) = \int d^2\boldsymbol{r}[(\partial_x^2 g)^2 + (\partial_x\partial_y g)^2  + (\partial_y\partial_x g)^2 + (\partial_y^2 g)^2]
\end{equation}

Figs. \label{fig:square_reconstruction} and \label{fig:meander_reconstruction} show the residuals and two components of the current density reconstructed from the DC flux images presented in the main text.

\begin{figure}
    \includegraphics[width=\columnwidth]{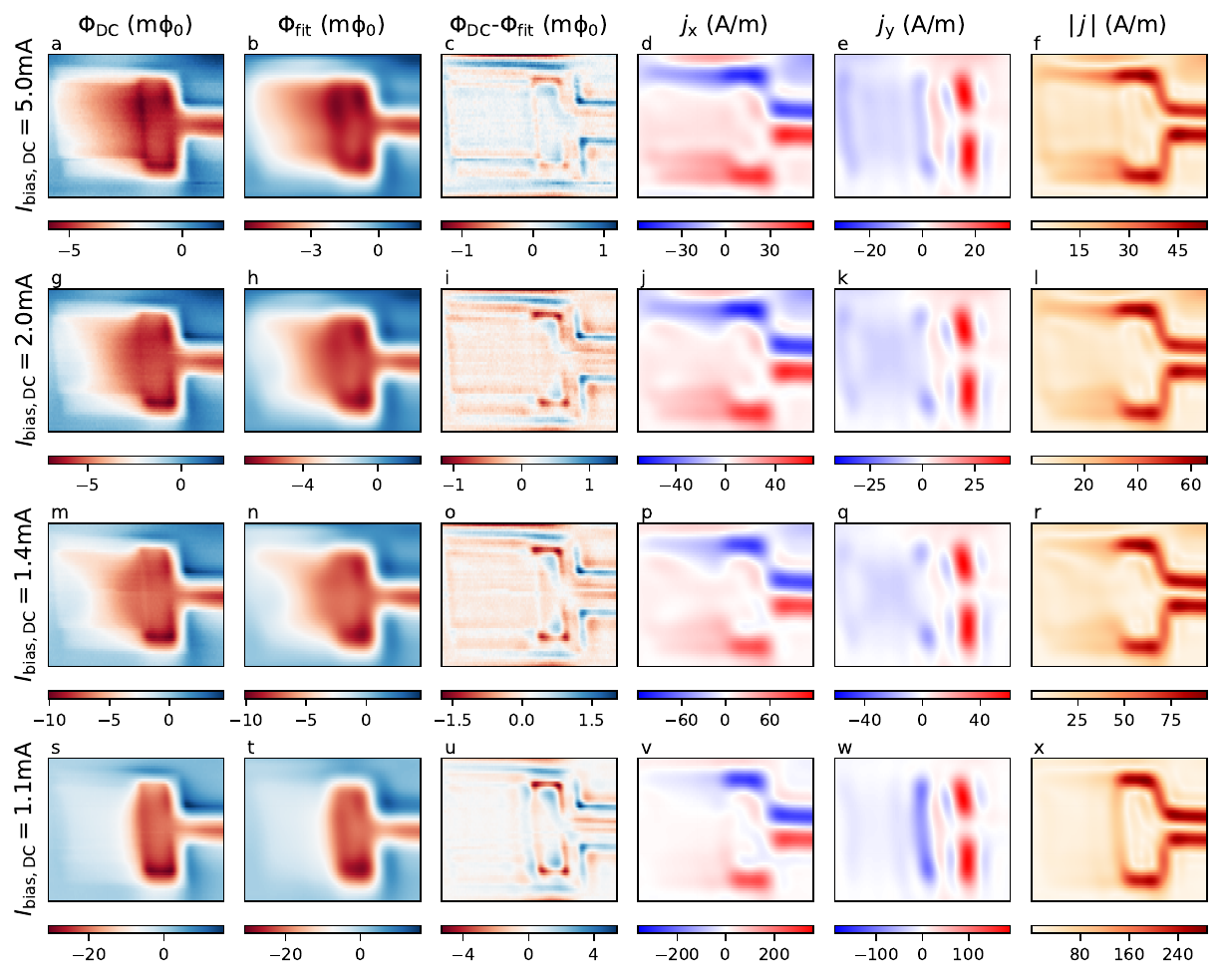}
    \caption{Components of the DC current reconstruction, where (a) and (f) are presented in the main text, for the square TES at different DC-bias points, 10\%, 7\%, 5\%, and 0\%\,$\Rn$ per row from top to bottom. 
    From left to right, columns show the measured DC-flux image $\Phi_\mathrm{DC}$, the reconstructed flux image $\Phi_\mathrm{fit}$, residuals between the measured and reconstructed flux images $\Phi_\mathrm{DC}$-$\Phi_\mathrm{fit}$, components of the in-plane current densities $j_x$ and $j_y$, and the magnitude of the current density $|j|$.
    }
    \label{fig:square_reconstruction}
\end{figure}

\begin{figure}
    \includegraphics[width=\columnwidth]{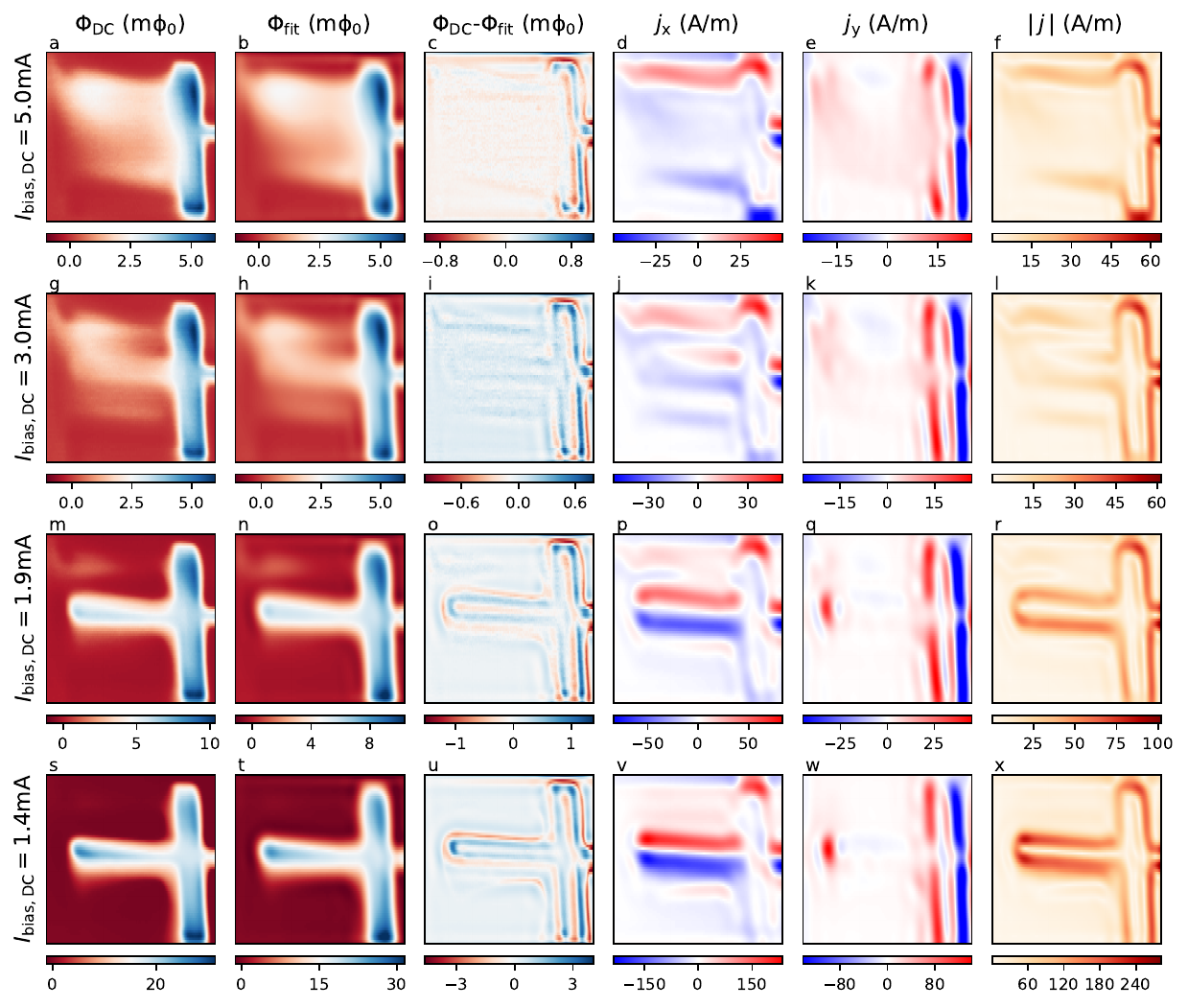}
    \caption{Components of the DC current reconstruction, where (a) and (f) are presented in the main text, for the meander TES at different DC-bias points, 68\%, 40\%, 11\%, and 0\%\,$\Rn$ per row from top to bottom. 
    From left to right, the columns show the measured DC-flux image $\Phi_\mathrm{DC}$, the reconstructed flux image $\Phi_\mathrm{fit}$, residuals between the measured and reconstructed flux images $\Phi_\mathrm{DC}$-$\Phi_\mathrm{fit}$, components of the in-plane current densities $j_x$ and $j_y$, and the magnitude of the current density $|j|$.
    }
    \label{fig:meander_reconstruction}
\end{figure}

\newpage
\clearpage
\newpage

\stepcounter{somecounter}
\section{Supplementary Section \thesomecounter: Current Distribution as a Function of Temperature}
In the main text, we present measurements of the TESs when they are voltage biased in their transition. The superconducting state changes as a function of bias revealed by our imaging of the diamagnetic response, and the transport current correspondingly reconfigures. Here, we set the DC-bias to zero and instead sweep temperature to gradually change the device from the superconducting to normal states. We apply only an AC-bias current of \SI{1}{\milli\ampere}. This is a rather large AC current, and it is very likely modifying the superconducting state across the values that the excitation samples its period. In this case we would measure the time-averaged behavior of the device throughout the period of the AC current. We choose the AC-current to be larger here in order to couple a reasonable amount of current through the TES when in the normal state to be able to image, as most of the current instead flows through the shunt resistor. 
\begin{figure}
    \includegraphics[width=.9\columnwidth]{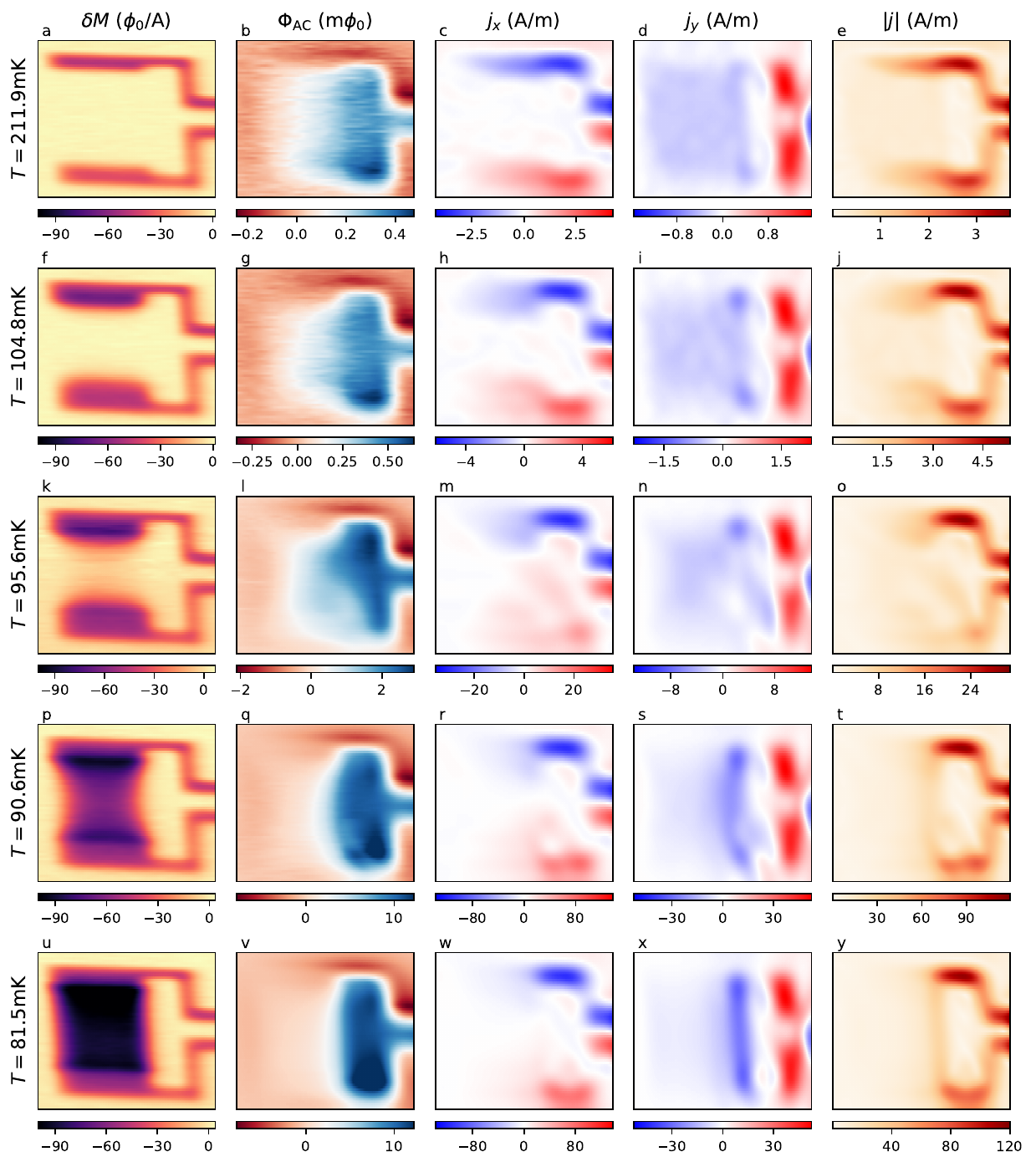}
    \caption{Imaging of the current distribution in the square TES as a function of temperature. Each row corresponds to a different temperature (leftmost axis), to tune the superconducting state of the device, as shown by AC susceptibility/diamagnetic response images of the device in the leftmost column. In the second to left column, we show the flux image associated with a \SI{1}{\milli\ampere} AC bias current. We perform a similar reconstruction of the current density from the flux image to the main text and show reconstructed in-plane current densities $j_x$ and $j_y$ and the magnitude of the current density $|j|$ in the rightmost three columns (left to right).}
    \label{fig:square_T}
\end{figure}

\begin{figure}
    \includegraphics[width=\columnwidth]{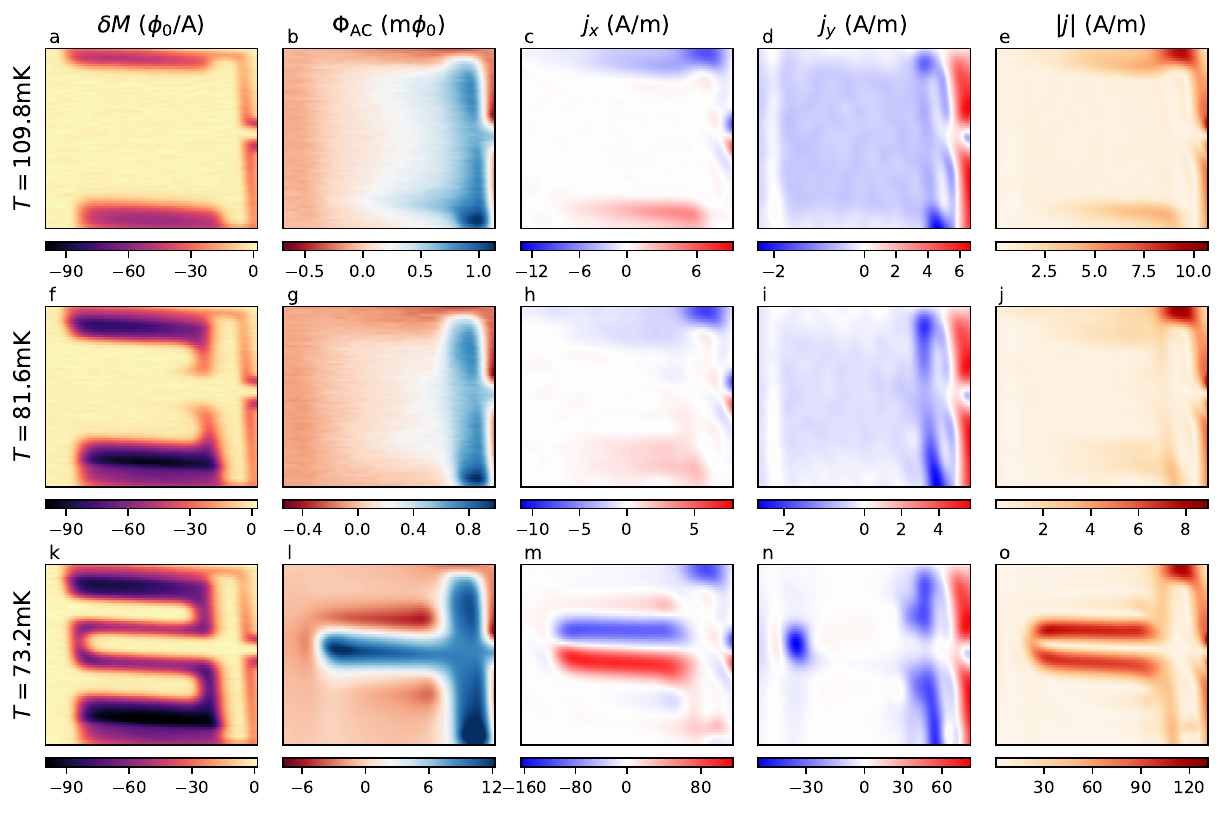}
    \caption{Imaging of the current distribution in the meander TES as a function of temperature. Each row corresponds to a different temperature (leftmost axis), to tune the superconducting state of the device, as shown by AC susceptibility/diamagnetic response images of the device in the leftmost column. In the second to left column, we show the flux image associated with a \SI{1}{\milli\ampere} AC bias current. We perform a similar reconstruction of the current density from the flux image to the main text and show reconstructed in-plane current densities $j_x$ and $j_y$ and the magnitude of the current density $|j|$ in the right three columns (left to right).
    }
    \label{fig:meander_T}
\end{figure}

\newpage
\newpage
\clearpage
\newpage

\stepcounter{somecounter}
\section{Supplementary Section \thesomecounter: London Modeling}
Superconductors placed in a magnetic field generate screening currents according to the London equation, which for a thin, 2D superconductor can be written as:
\begin{equation}
    \label{eq:London}
    \nabla \times \boldsymbol{j} = \boldsymbol{H}/\Lambda, 
\end{equation}
where $\Lambda$ is the Pearl length, and $\boldsymbol{j}$ is the thickness integrated current density. Transport currents flowing in a superconductor themselves produce magnetic fields, so the current distribution and its self-field must satisfy both Ampere's law and Eq.\ref{eq:London}. As a result, transport currents get pushed towards the edges of a device due to self-screening, while the Pearl length (or London penetration depth in 3D) determines how pronounced this effect is. In a uniform superconducting strip with uniform current injection, the current would be pushed to the two edges symmetrically. 

In the main text, we observe that in the TESs the current distribution is asymmetric with more current along the edge closer to the leads. Here, we solve the London equation approximating a fully superconducting square TES to understand whether the asymmetric lead geometry can explain the observed asymmetry in the current distribution \cite{BISHOPVANHORN2022}. This modeling is intended to be qualitative, and we therefore limit ourselves to the 2D case. The substantial thickness of the TES may limit the validity of this 2D approximation. A full 3D treatment would be desirable but is challenging, particularly because the superconducting properties may vary through the thickness of the bilayer.

\begin{figure}[htb]
    \includegraphics[width=.8\columnwidth]{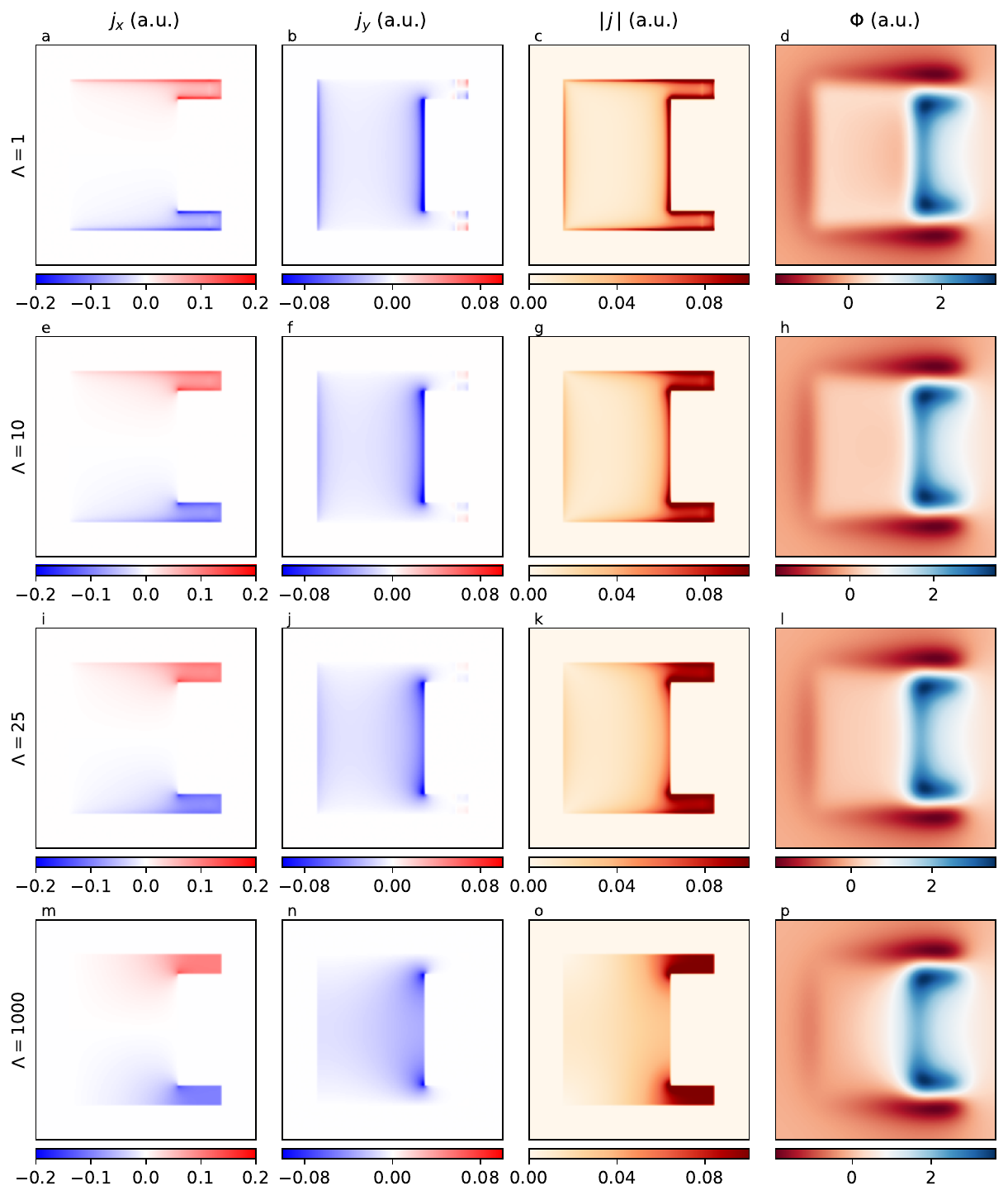}
    \caption{Modeled current densities and flux images when including London screening for the square TES when in the fully superconducting state. Each row corresponds to a different Pearl length and therefore screening strength. For small $\Lambda$ and large screening, we observe much of the transport current is squished to a narrow region at the right edge of the device, while there is another narrow segment of current along the left edge of the device which is not explicitly observed in our measurements. At very large $\Lambda$ and small screening, the current is injected in to the device and spreads out much more uniformly, in a similar fashion to an ohmic conductor. The reconstructed current in the square TES measured here looks to be similar to the case of $\Lambda\sim10$ where there is a squeezing of the current near the right edge of the device, but only a very weak current that flows along the narrow path at the left edge of the device.
    }
    \label{fig:london_img}

\end{figure}

\end{document}